\documentclass[12pt,preprint]{aastex}
\begin{document}


%\title{RHESSI OBSERVATIONS OF A COMPLEX LOOT STRUCTURE IN AN M7.6 FLARE ON 24 OCTOBER 2003: EVIDENCE FOR X-RAY ARCADE}
%\title{RHESSI OBSERVATIONS OF 2003 OCTOBER 24 LONG DURATION FLARE: CONSEQUENCES OF MAGNETIC RECONNECTION}
%\title{Two distinct phases of hard X-ray emissions in a solar eruptive flare}
\title{MAGNETIC RECONNECTION DURING THE TWO-PHASE EVOLUTION OF A SOLAR ERUPTIVE FLARE}


\author{Bhuwan Joshi\altaffilmark{1,2}, Astrid Veronig\altaffilmark{3}, K. -S. Cho\altaffilmark{1}, S. -C. Bong\altaffilmark{1}, B. V. Somov\altaffilmark{4}, Y. -J. Moon\altaffilmark{5},
Jeongwoo Lee\altaffilmark{6}, P. K. Manoharan\altaffilmark{7}, and Y. -H. Kim\altaffilmark{1}} 
\altaffiltext{1}{Korea Astronomy and Space Science Institute, Daejeon 305-348, Korea}
\altaffiltext{2}{Udaipur Solar Observatory, Physical Research Laboratory, Udaipur 313 001, India}
\altaffiltext{3}{IGAM/Institute of Physics, University of Graz, Universit$\ddot{a}$tsplatz 5, A-8010 Graz, Austria}
\altaffiltext{4}{Astronomical Institute, Moscow State University, Universitetskij Prospekt 13, Moscow 119992}
\altaffiltext{5}{School of Space Research, Kyung Hee University, Yongin 446-701, Korea}
\altaffiltext{6}{Physics Department, New Jersey Institute of Technology, 161 Warren Street, Newark, NJ 07102}
\altaffiltext{7}{Radio Astronomy Centre, Tata Institute of Fundamental Research, Udhagamandalam (Ooty) 643 001, India}
%\affil{Korea Astronomy and Space Science Institute, Daejeon, Korea}
\email{bhuwan@prl.res.in}

%\author{C. D. Biemesderfer\altaffilmark{4,5}}
%\affil{National Optical Astronomy Observatories, Tucson, AZ 85719}
%\email{aastex-help@aas.org}
%
%\and
%
%\author{R. J. Hanisch\altaffilmark{5}}
%\affil{Space Telescope Science Institute, Baltimore, MD 21218}


%\altaffiltext{1}{Visiting Astronomer, Cerro Tololo Inter-American Observatory.
%CTIO is operated by AURA, Inc.\ under contract to the National Science
%\altaffiltext{2}{Society of Fellows, Harvard University.}
%\altaffiltext{3}{present address: Center for Astrophysics,
%    60 Garden Street, Cambridge, MA 02138}
%\altaffiltext{4}{Visiting Programmer, Space Telescope Science Institute}
%\altaffiltext{5}{Patron, Alonso's Bar and Grill}


\begin{abstract}
We present a detailed multi-wavelength analysis and interpretation of the evolution of an M7.6 flare that occurred near the south-east limb on October 24, 2003. Preflare images at TRACE 195~{\AA} show that the bright and complex system of coronal loops already existed at the flaring site. The X-ray observations of the flare taken from the RHESSI spacecraft reveal two phases of the flare evolution. The first phase is characterized by the altitude decrease of the X-ray looptop (LT) source for  $\sim$11 minutes. Such a long duration of the descending LT source motion is reported for the first time. The EUV loops, located below the X-ray LT source, also undergo contraction with similar speed ($\sim$15~km~s$^{-1}$) in this interval.
During the second phase the two distinct hard X-ray footpoints (FP) sources are observed which correlate well with UV and H$\alpha$
flare ribbons. The X-ray LT source now exhibits upward motion as anticipated from the standard flare model. The RHESSI spectra during the first phase are soft and indicative of hot thermal emission from flaring loops with temperatures ~$T>25$~MK at the early stage.
On the other hand, the spectra at high energies ($\varepsilon \gtrsim$25~keV) follow hard power laws during the second phase ($\gamma = 2.6-2.8$). 
We show that the observed motion of the LT and FP sources can be understood as a consequence of three-dimensional magnetic reconnection at a separator in the corona. During the first phase of the flare, the reconnection releases an excess of magnetic energy related to the magnetic tensions generated before a flare by the shear flows in the photosphere. The relaxation of the associated magnetic shear in the corona by the reconnection process explains 
the descending motion of the LT source. During the second phase, the ordinary reconnection process dominates describing the energy release in terms of the standard model of large eruptive flares with increasing FP separation and upward motion of the LT source.
\end{abstract}


\section{Introduction}
The active Sun displays a stunning variety of transient energetic phenomena ranging
from the smallest microjets and microflares to the largest flares and coronal mass ejections (CMEs). It is generally accepted that the energy
released during flares and CMEs is stored in the corona prior to the event in the form of stressed or non-potential magnetic fields. We believe that
this stored free energy in the coronal magnetic fields is released explosively through the process of magnetic reconnection. Multi-wavelength
measurements of flares and CMEs made over the years have provided many valuable pieces of information about the response of coronal energy release
in the different atmospheric layers of the Sun. During an eruptive flare, the whole magnetic configuration of coronal loops crossing an inversion line
is disrupted and is followed by a newly rebuilt loop system from pre-flare to post-flare stages. The expansion of flare ribbons and growth of flare
loop system are the clearest observational findings which support the flare models involving magnetic reconnection \citep[for a review see][]{Priest02,Lin03}.

Yohkoh spacecraft significantly improved our overall understanding about the flare physics \citep{Hudson04}. Observations made
by Yohkoh confirmed the presence of hard X-ray sources at the footpoints of the loop system \citep{Sakao94} and detected a new hard 
X-ray source above the apex of the hot flaring loop observed in soft X-rays \citep{Masuda94,Somov05}. 
The above-the-looptop source 
is believed to be intimately connected to the primary energy release site and location of the
electron acceleration in the corona, again supporting the reconnection models. RHESSI spacecraft with superior observing capabilities
in X-ray energy levels, further refined our knowledge about the X-ray LT coronal sources \citep[see review by][]{Krucker08}. It detected a downward motion
of LT source during the initial phase of impulsive rise of hard X-ray flux \citep{Sui03,Sui04,Liu04,Ji06,Veronig06,Joshi07}. We still do not have a plausible explanation of this 
phenomenon. RHESSI observations have also revealed the existence of double coronal sources \citep{Sui03,Sui04,Sui05,Veronig06,Li07,Liu08}. The double coronal source has been interpreted as an indirect evidence for the 
formation and expansion of a large scale current sheet in the corona \citep{Sui03,Sui04,Sui05}.

In this paper we present a comprehensive multi-wavelength analysis of an M7.6 flare on 2003 October 24. A detailed X-ray 
imaging and spectroscopic analysis of RHESSI observations exhibit two phases of flare evolution in view 
of the X-ray energy release process. 
%The first stage is gradual and dominated by thermal emission while the second stage is 
%impulsive with strongly non-thermal characteristics.  
A significant and unusually long downward motion of the LT source 
characterizes the first phase. During the whole second phase we observe footpoints sources and upward expansion of the LT source, as it is commonly observed in solar flares. The event was associated with a fast CME. We compare flare observations at different 
X-ray wavelengths, and also with other chromospheric and coronal images with the aim to study the consequences of the magnetic
reconnection process at various atmospheric layers.
In section 2, we describe the observational characteristics of the event. We interpret our observational
findings in the final section of the paper.
%
%
%In this paper, we concentrate on the detailed RHESSI observations of a well-observed M7.6 flare on 2003 October 24. 
%The event was associated with a fast CME.
%We compare flare observations at different X-ray wavelengths, and also with other chromospheric and coronal images with 
%the aim to study the consequences of the magnetic reconnection
%process at various atmospheric layers. 

\section{Observations and Results}
%Figure \ref{goes} shows the soft X-ray (SXR) flux recorded by the GOES satellite in the 0.5--4 and 1--8~{\AA} wavelength bands between 02:00 to 03:40 UT.
%GOES measurements reveal this flare to be a typical long duration event lasting from 02:15 to about 05:00 UT with a maximum around 02:53 UT.
%The Reuven Ramaty High Energy Solar Spectroscopic Imager (RHESSI; Lin et al. 2002)  made a complete coverage of this long duration event.
%The Transition Region and Coronal Explorer (TRACE; Handy et al. 1999) had obtained image sequences in 195~{\AA} and 1600~{\AA} wavelengths of the
%activity site during various phase of the flare evolution. In the following subsections we analyze these observations in detail.

\subsection{X-ray time profiles: two phase evolution}

The soft X-ray fluxes recorded by the GOES satellite in the 0.5--4 and 1--8~{\AA} wavelength bands reveal this flare to be a typical long
duration event lasting from 02:15 to about 05:00 UT. The GOES flux in the 0.5--4~{\AA} channel attained the first peak at 02:40 UT (Fig. \ref{goes_rhessi}). 
The flux then shows a decreasing trend and then further increases after 02:44 UT. The overall maximum is achieved in both GOES channels at 02:53 UT.

The Reuven Ramaty High Energy Solar Spectroscopic Imager \citep[RHESSI;][]{Lin02}  made a complete coverage of this long duration event.
However, X-ray counts were contaminated by counts produced by radiation belt particles hitting the RHESSI detectors from all the directions.
Particle events mainly affect the X-ray count rate at high energies ($\geq$ 40 keV). In order to obtain the X-ray fluxes
free from particle contamination we adopted the method described in \cite{Liu09}.
%Liu et al. (2009a). 
For the background estimation we selected a 
non-flare interval 04:30:30--04:33:00 on 2003 October 24, when the spacecraft was at approximately same geomagnetic location
during its next orbit. The corrected X-ray count rates averaged over front detectors 1, 3--6, 8, and 9 in the 6--12, 12--25, 25--50, 50--100, and
100--300 keV energy bands are shown in Figure \ref{goes_rhessi}.

The inspection of GOES and RHESSI ($\leq$ 25 keV) time profiles shows two phases of the flare evolution: 
a first phase between 02:24 and 02:44 UT, followed by a second phase. 
However, the X-ray light curves at high energy bands ($\geq$ 25 keV) reveal that the emission during the second phase is much stronger 
than the first phase. Here it is relevant to mention that the X-ray and EUV images further confirm two phase distinction in 
terms of the morphological evolution of the flare (section \ref{section_xray_euv}).

It is noteworthy that the LASCO instrument onboard SOHO \citep{Brueckner95} detected a fast CME associated 
with this flare event. The height-time plot
available at SOHO--LASCO--CME catalogue\footnote{http://cdaw.gsfc.nasa.gov/CME$\_$list/} shows the mean propagation speed of CME in LASCO field of view
to be about 1055~km~s$^{-1}$. The CME first appeared in the LASCO C2 coronagraph at 02:54 UT at a radial distance of $\sim$2.7~R$_\odot$.
The extrapolation of CME height backward in time suggests its association with the first phase of the flare evolution. 

\subsection{X-ray, (E)UV and H$\alpha$ imaging}
\label{section_xray_euv}
The RHESSI images have been reconstructed with the CLEAN algorithm with the natural weighing scheme using front detector segments 3 to 8 in different
energy bands, namely, 6--12, 10--15, 25--50, 50--100, and 100-300 keV \citep{Hurford02}. We compare RHESSI measurements with TRACE images in 195 and 1600~{\AA}
wavelengths. The TRACE 1600~{\AA} channel is sensitive to plasma in the temperature range between (4--10)$\times$10$^{3}$ K. Emissions at this
wavelength originate mainly from the chromosphere and provide valuable informations about the chromospheric response to the coronal energy release.
The TRACE 195~{\AA} filter is mainly sensitive to plasmas at a temperature around 1.5 MK (Fe XII) but during flares it may also contain significant contributions of plasmas at temperatures around 15--20 MK \citep[due to an Fe XXIV line;][]{Handy99}.

%The TRACE 195~{\AA} channel is sensitive to higher temperature %[(5.0--20)$\times$10$^{5}$ and (1.1--2.6)$\times$10$^{7}$ K] plasma and 
%show relatively low temperature coronal structures which develop during the %flare evolution \citep{Handy99}.

It is well known that TRACE images do not have accurate absolute pointing. However, the pointing information of RHESSI and SOHO is believed to be accurate. Therefore we corrected TRACE pointing by a cross correlation alignment between a TRACE and a SOHO/EIT image observed at 02:23:30 and 02:24:38 UT, respectively, in the 195~{\AA} channel. For the cross correlation, we used the Solar SoftWare (SSW) routine, trace\_mdi\_align, developed by T. Metcalf \citep[see also][]{Metcalf03}.

First, we discuss the flare characteristics during the first phase of evolution (i.e., between 02:24 and 02:44 UT) prominent in GOES profile and
RHESSI measurements at low energies ($\lesssim$ 25 keV). 
%The sequence of RHESSI images
% in 6--12 and 12--25 keV 
%at low energies ($\leq$ 25~keV) show 
RHESSI images reconstructed in various energy bands below 50 keV show a bright LT source which develops right at the flare onset. The LT source then shows shift to lower altitudes. 
%Images in both the energy bands reveal similar morphological and temporal evolution of looptop source.
In Fig. \ref{rhessi_first_ph}, four representative 6--12 keV images during this period are shown. At the flare onset, a LT source and associated
coronal loop are seen. The LT source appeared above the limb while the loop system extends to the disk with northern leg
apparently longer than the southern one (top left panel of Fig. \ref{rhessi_first_ph}). In the later stages, the loop system shrinks and 
the LT structure continues to become more compact and brighter.
 
%The examination of TRACE 195~{\AA} images reveals that 
A bright and complex system of parallel loops already pre-existed and was observable in TRACE 195~{\AA} images at the  
activity site before the flare onset (first panel of 
Fig. \ref{trace195_series}). RHESSI~LT source, which appeared as early as $\sim$02:24 UT, is located above 
the northern side of the TRACE loop system (Fig. \ref{trace195_hessi}). 
Intense brightening occurred at the top of TRACE loops from 02:29 UT 
onward and the loops start to shrink (Fig. \ref{trace195_series}). 

In Fig. \ref{trace_rhessi_LT} we present the altitude evolution of
the LT source derived from RHESSI images in three energy bands, namely 6--12, 12--25 and 25--50 keV, and TRACE 195~{\AA} images. Here, the LT altitude is defined as the distance, along the 
main axis of motion, between the centroid of LT and the center of the line between the two FPs seen in RHESSI 50--100 keV energy band image at 02:51:00 UT. The axis of motion is determined by fitting a straight line to the centroids of the LT source and is inclined at an angle of 25$^{o}$ northward from the radial X-axis. For RHESSI images, we have taken the centroid of emission of all pixels above 85\% of the peak flux. The TRACE LT is defined by selecting a region near the top of loop with emission above 90\% of the peak flux of each image. We find that the RHESSI LT source observed at 6--12, 12--25 and 25--50 keV energy bands 
%located at higher altitude than the LT sources at 6--12 and 12--25 keV 
%energy bands. The LT sources observed at 6--12 and 12--25 keV energy bands 
are almost co-spatial during the whole first phase and descend with similar velocities between 02:25 and 02:35 UT ($\sim$15~km~s$^{-1}$). The TRACE~LT source, located below the X-ray LT source, also shows similar velocity of downward motion in this interval. RHESSI and TRACE images provide a consistent picture of the loop shrinkage during the first phase. 

%We find that the RHESSI~LT source observed at 6--12 and 12--25 keV energy band images are co-spatial during the course of downward motion between 02:25 and 02:35 UT and descend with a mean velocity of 15 and 14~km~s$^{-1}$ respectively. The TRACE~LT source, located below the X-ray LT source, also shows similar velocity of downward motion in this interval (16~km~s$^{-1}$). 

RHESSI images during the second phase of the flare (i.e., between 02:44 to 03:10 UT) show the expansion of flare loop at low energies ($\lesssim$25~ keV), and the origin and evolution of the hard X-ray sources on the solar disk at high energies ($\gtrsim 25$~keV). Figure \ref{rhessi_second_ph} presents a few representative images in 10--15 and 50--100 keV energy bands. The X-ray light curve above 25 keV shows peaks at the time of these images (Fig. \ref{goes_rhessi}). A hard X-ray source appeared around the peak at $\sim$02:45 UT. Another source developed southward to the first one at 02:48 UT, the time of flare maximum observed in RHESSI light curves above 25 keV. 
These two sources lie over the two bright flare ribbons observed at 1600~{\AA} TRACE images (Fig. \ref{trace1600_hessi}). 
Therefore, we interpret these two sources as the northern and southern FPs of a loop system, the LT of which is clearly visible in RHESSI images at low energies (Figs. \ref{rhessi_second_ph} and \ref{trace1600_hessi}). We note that most of the time the northern FP remains brighter than the southern one. However, the intensity of the southern FP increases slowly with time and near the end of the second phase both the FPs are of similar brightness.

We note that the evolution of the LT source after 02:35 UT is rather complicated. Examination of both 6--12 and 12--25 keV images shows that
the height of LT source does not change for the next few minutes before showing the commonly observed upward expansion (Fig \ref{trace_rhessi_LT}).
The X-ray LT source at 12-25 keV starts moving upward $\sim$3 minutes earlier than the 6-12 keV source. 
%Here we should bear in mind that 12-25 keV source observes hotter plasma %than 6-12 keV source.
We also notice that during the upward expansion, the LT source at 12--25 keV 
is located at higher altitudes than the LT source in 6--12 keV (i.e. the emission at higher X-ray energies originates from higher altitude). 
For upward motion, we obtained mean velocities of 22 and 17~km~s$^{-1}$ for
6--12 (between 02:49--03:00 UT) and 12--25 keV (between 02:46--03:00) sources respectively.
Another interesting feature during the upward motion of the X-ray LT source is the increase in the brightness at the top of a nearby loop system 
(seen in TRACE 195~{\AA} images; see image at 02:58:59 UT in Fig. \ref{trace195_hessi}) situated at the southern side of the main LT source. 
In the later stages, the brightness of southern LT source increases and start dominating over the northern main LT source. 
The TRACE~195~{\AA} LT source with bright, diffuse emission exhibits little downward motion between 02:35 and 02:50 UT
(Figs. \ref{trace195_series} and \ref{trace_rhessi_LT}) with a mean velocity of 4~km~s$^{-1}$. 
In the later stages ($>$ 02:54 UT), the bright emission seen in TRACE~195~{\AA} images is more structured and spread along the loop, below which 
a dark (and cool) system of post-flare loops is visible (Figure \ref{trace195_series}). The loop system shows an upward expansion with 
a mean velocity of 4~km~s$^{-1}$.

Here it is important to discuss a few observational features seen in the X-ray light curve and images for 25--50 keV energy bands in detail (Figs. \ref{goes_rhessi} and \ref{rhessi_25-50}). There is a sharp rise in the X-ray flux from 02:27 UT which attains a maximum level at 02:30 UT. The flux maintains nearly a constant level for the next few minutes and shows a gradual decline till $\sim$02:43 UT. The images during this interval show a descending LT source, similar to the LT source observed at energies below 25 keV. However, for a brief period of time, 02:30--02:31~UT, two bright and distinct FP sources are observed in addition to the LT source (Figure \ref{rhessi_25-50}b). The 25--50 keV images during the second phase show only FP sources and its evolution is similar to the FP sources seen at higher energies ($\geq$ 50~keV). 
In Fig. \ref{trace_rhessi_LT} (top panel) we also show the separation of the two X-ray FPs reconstructed at 25--50~keV energies. The two distinct FPs, observed between 02:48 and 02:57 UT, separate from each other with average speed of $\sim$2~km~s$^{-1}$. We also find that the speed of separation increases at later stages ($\sim$7~km~s$^{-1}$ between 02:52 and 02:57 UT). We further notice that the separation between the two FPs in the first phase (which is observed only in one image at 02:30--02:31) is larger than the FP separation observed in the second phase (see Figs \ref{rhessi_25-50}b and \ref{rhessi_25-50}e). 

%This means that the bases of the magnetic loops are moving towards each %other during the first phase. The images also reveal that the FPs are mainly %moving along the Y-direction; northern FP moves southward while the southern %FP moves northward. The FPs come closer to each other along the Y-direction %by $\sim$20\arcsec~between 02:30 and 02:52 UT.  On the other hand, the %movement of FPs along the X-direction is very small during this period; the %northern FP shows a shift of $\sim$6\arcsec~westward while southern FP does %not show any significant shift along the X-direction. In contrast, FPs do %not show significant motions during the second phase (see Figs %\ref{rhessi_25-50}e and \ref{rhessi_25-50}f). 
%There is a possibility that FPs are moving along the X-direction during this %phase, however, because of the projection effects near the limb it is hard %to detect such motions.

In the bottom panel of Figure \ref{trace_rhessi_LT}, we have shown the temporal evolution of plasma temperature derived from GOES 12 observations.
We find that the plasma temperature is higher during the first phase and attains the maximum value at $\sim$02:34 UT. Further we find that
the decrease of the LT source altitude is anti-correlated with plasma temperature evolution during the first phase (between 02:24 and 02:35 UT). After 02:34 UT
plasma temperature decreases rapidly till $\sim$02:44 UT. In the beginning of the second phase the temperature remains nearly constants for several minutes
(between $\sim$02:47 and $\sim$02:53 UT) while it decreases very slowly in the later stages.

%The flare associated bright plasma structures observed in TRACE 195~{\AA} %images during the first and second phases 
%need further attention. Throughout the first phase, the intense LT source is %seen in TRACE~195~{\AA} images below X-ray LT source,
%is co-spatial with X-ray looptop source observed in 6-25 keV energies and shows similar altitude evolution. This diffuse emission
%is believed to be originated from hot 15-20 MK plasma in response to the %presence of Fe XXIV (192~{\AA}) resonance line which lies within the 
%TRACE 195~{\AA} passband (Warren \& Reeves 2001). Warren \& Reeves (2001) %and Gallagher et al. (2002) also noticed similar flare associated
%high-temperature plasma structures. On the other hand, the bright, slowly %expanding loop system 
%seen in TRACE~195~{\AA} images during the later stages ($>$ 02:54 UT) lies %below the upward moving RHESSI X-ray looptop source (Fig. %\ref{trace195_hessi}).
%This suggests that now TRACE at 195~{\AA} observes post-flare loop system %and bright, structured emission along the loops indicates
%low temperature plasma structures (1.5 MK) dominating the TRACE~195~{\AA} %filter response. 
%We also notice that between 02:50 and 02:54 UT the TRACE LT source shows %fast downward motion
%with mean velocity of 19~km~s$^{-1}$, which is quite different from the %behaviour of RHESSI LT source (Fig. \ref{trace_rhessi_LT}). 
%This likely reflects the fact that during this interval in TRACE~195~{\AA} %images,
%we switch from 15-20 MK flare plasma to 1.5 MK post-flare loop system.

The TRACE images at 1600~{\AA} and H$\alpha$ filtergrams taken from ARIES Solar Tower Telescope, India \citep{Pant06} during the 
decline phase (after 02:55 UT) reveal clear post-flare loop configuration 
that connects the flare ribbons (Figs. \ref{trace1600_hessi} and \ref{halpha_hessi}). 
%The loop system grows and looptop gets intensified after 02:59 UT. 
%The decay phase of the flare was observed in H$\alpha$ from ARIES Solar Tower Telescope, Nainital, India (Pant 2006). Figure \ref{halpha_hessi}
%presents a few selected H$\alpha$ filtergrams which reveal two parallel flare ribbons connected by a bright loop system. 
%We find that the morphological details
%of flaring region in H$\alpha$ filtergrams and TRACE 1600~{\AA} images are very similar. 
We further notice that the H$\alpha$ loops are seen in emission against the solar disk which is not a commonly observed phenomenon.
%which is rather unusual and indicative of high densities 
%in the post flare loops {\bf \citep[in excess of $10^{12}$~$cm^{-3}$; %see][]{Heinzel87}}.

\subsection{RHESSI X-ray spectroscopy}

Figure~\ref{rhessi_spec} shows spatially integrated, background subtracted RHESSI spectra derived 
during nine time intervals of the flare together with the applied spectral fits. 
Each spectrum was accumulated over 80~s and derived with 1 keV energy bins from 10 to 200~keV using all RHESSI front detector segments except 2 and 7 (which have lower energy resolution and high threshold energies, respectively) and deconvolved with the full detector response matrix, i.e., including off-diagonal elements \citep [see][]{Smith02}. 
Due to the high count rates at low energies, the effect of pile-up may be significant during this event, 
i.e.\ two (or more) low-energy photons arrive almost simultaneously
at one detector and are recorded as a single photon with
energy equal to the sum of the individual photon energies \citep{Smith02}.
Detector live time is the simplest indicator for a quantitative understanding of the pile-up effect. We obtained live time averaged over the seven detectors being used. The spectra shown in top row of Fig.~\ref{rhessi_spec} were derived during the first phase of the flare. The average live time for these three time intervals was 92\%, 86\% and 97\% respectively. We find that for the third interval the pile-up effect is smallest (detector live times are 97\%) because it is in the A3 attenuator state. During the second phase we obtained spectra for six time intervals (middle and bottom rows of Fig.~\ref{rhessi_spec}) with average live time of 83\%, 81\%, 81\%, 81\%, 83\%, and 91\% respectively. We note that the spectra derived during the interval 02:48:00--02:49:20 covers the impulsive phase for which the average live time was 81\%. In general, the live time values are relatively lower and indicate moderate pile-up severity. For example the average live time for the M1.7 flare on 2003 November 13 during the impulsive phase was ranging from 96\% to 89\% \citep{Liu06}. We applied the pile-up correction implemented in the RHESSI software. However, we note that during the early phase (Fig.~\ref{rhessi_spec}, top row) where the spectra are very steep, this correction may not be fully satisfactory. 

Spectral fits were obtained using a forward-fitting technique
for which the functional form of the incident photon flux spectra
is assumed for a parametric model of the source. Specifically,
we used the bremsstrahlung spectrum of an isothermal plasma and a power-law 
function with a turn-over at low X-ray energies.
The spectra during the first phase of the flare (top row of Fig. \ref{rhessi_spec}) %are dominated by 
suggest hot thermal emission with 
temperatures~$T >25$~MK already at the very start. At X-ray energies $\varepsilon \gtrsim 20$~keV,
the spectrum shows a steep power-law, with photon spectral index $\gamma \gtrsim 7$. At this first phase of the flare, RHESSI images show emission mostly from the flaring loops, concentrated toward the loop tops. Only during one interval at 
02:30--02:31~UT we briefly see HXR emission from footpoints (in 25--50~keV RHESSI images, Fig \ref{rhessi_25-50}). At this time
also the spectral index is harder with $\gamma = 5.7$ (see middle panel of top row in Fig.~\ref{rhessi_spec}).

During the second phase of the flare, the spectra show quite different characteristics (see middle and bottom rows of Fig.~\ref{rhessi_spec}). Emission up to $>$200~keV from RHESSI footpoints is detectable above the background, and the spectrum at high energies follows hard power laws (with photon spectral index $\gamma$ in the range 2.5 to 2.8 during the individual peaks). We note that the thermal emission is not increasing any further in this second flare phase but temperature as well as emission measure reached their peaks already during the first phase. %(see Fig.~\ref{rhessi_spec}).

\section{Discussion and Interpretation}

\subsection{The first phase}
Preflare images of the active region taken by TRACE at 195~{\AA} show that a bright and complex system of coronal loops already existed in the flaring region. The X-ray and EUV observations during the first phase of flare evolution reveal many important observational findings.
The X-ray LT source appeared right at the flare onset and starts moving to lower altitudes. The LT source is characterized by unusually long
duration of altitude decrease (from 02:24 to 02:35 UT). 
About 5 minutes after the appearance of X-ray LT source, an intense and diffuse brightening is observed at the top of TRACE~195~{\AA} loops, located below RHESSI LT source. \cite{Warren01} and \cite{Gallagher02} reported very similar bright and diffuse emission from TRACE~195~{\AA} loops associated with X-class flares, which was interpreted as being due to the Fe XXIV contribution of plasma at a temperature of 15-20 MK. 
\cite{Phillips05} further confirmed that the Fe XXIV line emission is the dominant contributor to emission in the TRACE~195~{\AA} channel for flare associated high temperature features (such as hot LT emission). These bright EUV loops showed a large scale contraction. The descending motion of X-ray LT source ceased at 02:35 UT while the EUV coronal loops continued to contract several minutes later, which we interpret as an effect of plasma cooling (from X-rays to EUV). The speeds of downward motion for EUV loops and X-ray LT source are comparable during the descending phase of X-ray LT source. Another noticeable feature in 25--50 keV X-ray images is the appearance of two bright FP sources in this phase for a short interval (02:30:00-02:31:00). 

The downward motion of LT source in the early phase of flare evolution has been established by many RHESSI observations {\citep{Krucker03,Sui03,Sui04,Liu04,Ji06,Veronig06,Joshi07,Liu08}.
Some of the previously studied events also show energy dependent structure of the LT source during the phase of altitude decrease, i.e. 
the higher energy sources were located higher in the corona and showed a faster speed of downward motion than lower energy sources
(Sui et al. 2004; Veronig et al. 2006; Liu et al. 2008). 
%{\bf In our event, we also find that the LT source at 25--50 keV is located %higher than the LT sources at 6--12 and 12--25 keV energy bands. However, %the LT sources in all the three energy bands show similar speeds of downward %motion.}
In our event, however, the difference of LT altitude between the 
sources observed at low and high energy
bands is not significant during this stage. 
Further the LT altitude decrease is observed in RHESSI observations for $\sim$11 minutes, 
which is significantly longer than reported previously.
The contraction of flaring EUV coronal loops has been reported by \cite{Li06}, \cite{Ji07}, and \cite{LiuR09}. \cite {Li05} reported the shrinkage of flare loops from radio observations at 34 GHz. 

Finally we note that the duration of downward motion of the LT source is marked by high plasma temperature. RHESSI observations indicate that the plasma
temperature 
%from 21$-$27 MK during this interval. We further notice that temperature
reached to high values ($\sim$27 MK) at the early stage.
The temperature determined from GOES channels is lower (maximum temperature $\approx$ 20 MK) and shows gradual rise with the LT shrinkage.
Large difference between temperature computed from RHESSI and GOES measurements could be due to different sensitivity and response of
these two instruments. This is also consistent with the fact that the observed plasma is multi-temperature. These results suggest a possible connection between LT heating and loop contraction.

\subsection{The second phase}

The flare evolution during the second phase 
%(i.e., between 02:44 and 03:10 UT) 
shows impulsive hard X-ray emission  
at high energies. During this interval various
observational signatures in H$\alpha$, (E)UV and hard X-ray wavelengths can be well understood by the standard CSHKP 
\citep{Carmichael64,Sturrock66,Hirayama74,Kopp76} model of eruptive
flares. These observational features include: appearance of two parallel flare ribbons in TRACE 1600~{\AA} and H$\alpha$ images;
two hard X-ray FPs with increasing separation, one lying on each flare ribbons; growth of X-ray LT height;
formation of relatively cool loop system below X-ray LT source at UV and H$\alpha$ wavelengths connecting 
the flare ribbons as well as post flare loop configuration
observed in the later stages of gradual decline of X-ray intensity. 
RHESSI and GOES measurements suggest that there is no further 
increase in temperature during the second phase. 
%At high energies the RHESSI spectra follow hard power laws ($\gamma$ $<$ 3) %which indicate the emission is dominated by energy released into %acceleration of electrons.
The flat X-ray spectra at high energies ($\gamma$ $<$ 3) obtained during the second phase indicate strong non-thermal emission, %, i.e., the emission is 
dominated by energy released into acceleration of electrons.

The beginning of the second phase is marked by the appearance of a HXR FP source which lie on the northern UV flare ribbon. At the flare maximum, second FP source developed on the southern flare ribbon.
However, the intensity of the southern FP increases continuously 
after the flare maximum and eventually both the FPs became of comparable brightness (Fig. \ref{rhessi_second_ph}). 
The consistency between the timings of peaks in light
curves and morphological 
evolution of FP sources can be understood in terms of thick-target model \citep{Brown71,Hudson72,Syrovatskii72}
in which the X-ray production at the FPs of the loop system takes place when high energy electrons, accelerated in the 
reconnection region, come along the guiding magnetic
field lines and slam the denser transition region and chromospheric layers producing hard X-ray emission via nonthermal bremsstrahlung
produced by fast electrons scattered off ions. 
The evolution of FP sources from
asymmetric to symmetric brightness can be seen as an evidence for different injection conditions of high energy electrons along the two 
legs of the loop system \citep{Siarkowski04} which could be mainly because of the highly complicated systems of loops.  
We find that the upward motion of 12--25~keV~LT source starts at the flare maximum. This upward growth
indicates progression of magnetic reconnection in the higher coronal loops rooted successively further apart from the magnetic
inversion line. The beginning of the ascending motion of X-ray LT source at the overall maximum of HXR time profile has been reported in several flares observed by RHESSI \citep{Sui03,Sui04,Liu04,Veronig06,Liu08}. On the other hand the lower loops eventually cool down (and fade away) and begin to be visible in low energies in UV and H$\alpha$ lines.
The structure of flaring region seen in H$\alpha$ filtergrams at this stage also reveal
a loop system connecting two H$\alpha$ flare ribbons and is very similar to the 1600~{\AA} images. We find that intense emission is produced from
the top of the H$\alpha$ loops. The appearance of bright looptop source 
in H$\alpha$ emission against the solar disk was also observed in the LT event studied in \cite{Veronig06} and \cite{Joshi07}. It is rather an unusual feature and considered as an evidence for high pressure, i.e. high density ($n \gtrsim 10^{12}$~$cm^{-3}$) in the observed postflare loops \citep{Heinzel87}.

\subsection{Overall picture of the flare evolution}

In order to interpret our observational results, we have to recall
some general properties of energy accumulation before a solar flare
and energy release during a flare.
On the basis of data obtained by the Hard X-ray Telescope
{\em HXT\/} on the satellite {\em Yohkoh\/},
\cite{Somov02,Somov03}
suggested that the large-scale structure and dynamics of
two-ribbon flares, as seen in hard X-rays (HXR), can be explained in
terms of the three-dimensional reconnection at a separator in the
corona.
More specifically, they suggested that, before a large two ribbon flare
%with observed significant decrease of a distance between the HXR
%conjugate foot\-points (FPs),
the bases of magnetic field lines are
moved by the large-scale photospheric flows of two types.
First, the converging flows, i.e., the flows directed to the
photospheric neutral line ($ PNL $), create the pre-flare current
layers in the corona and provided an excess of magnetic energy
sufficient to produce a flare.
Second, the shear flows, which are parallel to the $ PNL $, increase
the length of field lines in the corona and, therefore, produce an
excess of energy too.

During a flare, both excesses of magnetic energy are released completely or partially. In this way, the model describes two kinds of apparent motions of
the HXR kernels. One involves the increasing distance between the chromospheric flare ribbons (and associated HXR kernels)
%One is an increase of the distance between the chromospheric flare ribbons in which the HXR kernels appear. 
which results from reconnection in a coronal current layer. The other involves the approaching kernels and is associated 
%The second effect is a decrease of the distance between the kernels. They move to each other as a result of 
with the relaxation of magnetic tensions generated by the photospheric shear flows before a flare. We call the latter process {\em shear relaxation\/} \citep[for more details][] {Somov07}.

Let us clarify the general situation in terms of a specific model
called ``rainbow reconnection'' \citep{Somov86}.
%(So\-mov~1985, 1986).
Figure~\ref{mag_photo}a
%
%%% Fig. 13a %%%
%
shows the simplest presentation of a bipolar distribution of the
vertical component~$ B_z $ of magnetic field in the photosphere.
The neutral line~$ PNL $ divides the region into two zones with
field polarities~$ N $ and $ S $.
This region may be deformed by horizontal photospheric flows with
the velocity field~$ {\bf V} $ in such a way that the $ PNL $
gradually acquires the shape of the letter $ S $ as shown in
Figure~\ref{mag_photo}b.
%
%%% Fig. 13b %%%
%
%
Beginning at a critical value of the curvature of the $ PNL $ \citep[see][]{Gorbachev88} in the magnetic field above the photosphere, calculated in the potential approximation, there appears a topologically
featured field line, a separator, as illustrated by Figure~\ref{rainbow}.
%
%%% Fig. 14 %%%
%
The separator~$ X $ is located above the $ PNL $ like a rainbow above
a river which makes a bend.
The separator is the place where a reconnecting current layer can
be created, and therefore the pre\-flare energy accumulation can begin.

%
%%%%%%%%%%%%%%%%%%%%%%%%%%%%%%%%%%%%%%%%%%%%%%%%%%%%%%%%%%%%%%%%%%%%%%%%
%
On the other hand, Figure~\ref{mag_photo}c
%
%%% Fig. 13 c %%%
%
shows that a large-scale vortex-type flow generates two components
of the velocity field in the photosphere: the velocity
components~$ {\bf V}_{\|} $
and $ {\bf V}_{\bot} $ are parallel and perpendicular
to the $ PNL $.
The first component provides a shear of field lines (magnetic shear)
above the $ PNL $.
The shear flow before a flare creates the longer magnetic loops which,
being reconnected mainly during the first phase of a flare, provide
the bright HXR kernels with a large foot\-point separation
(Figure~\ref{flare_cartoon}).
%
%%% Fig. 15 %%%
%
The second component $ {\bf V}_{\bot} $ of the velocity field in the
photosphere tends to drive reconnection in the corona.
To demonstrate the basic physics in the simplest way, we
consider only a central region~$ C $ in the vicinity of
the $ S $-shaped neutral line~$ PNL $ in
Figure~\ref{mag_photo}b.
%
%%% Fig. 13 b %%%
%
Here we put the $ y $-direction along the
$ PNL $; the separator is
nearly parallel to $ PNL $ as was shown in
Figure~\ref{rainbow}.

Being reconnected at the separator, each magnetic field line as well
as each tube of magnetic flux~$ f $ is initially accelerated to
high velocity
($ \stackrel{ > }{ _\sim } 1000 $ km/s) inside a super-hot
turbulent-current layer \citep[SHTCL;][Chapter~6]{Somov07}. 
Each tube of reconnected field lines, being frozen into super-hot
($ T_{e} \stackrel{ > }{ _\sim } 10^8 $~K)
plasma, moves out of the SHTCL;
in the down\-flow it forms a magnetic loop with properties of a
collapsing magnetic trap \citep[][Chapter~7]{Somov97,Somov07}.
A collapsing magnetic trap model has been
first applied to one of the RHESSI events showing a descending LT source
by \cite {Veronig06}. The longitudinal and transverse sizes of the trap decrease, causing the trapped particles to acquire an additional energy
under Fermi and betatron acceleration respectively \citep{Somov03a}.
The energy distribution of trapped electrons and their HXR emission
can be calculated as a function of the trap length and its
thickness \citep{Bogachev07}.

%
%%%%%%%%%%%%%%%%%%%%%%%%%%%%%%%%%%%%%%%%%%%%%%%%%%%%%%%%%%%%%%%%%%%%%%%%
%
If the electrons injected into the trap from the SHTCL have a
power-law energy distribution, then their spectrum remains a
power-law one throughout the acceleration process for both the
Fermi and betatron mechanisms.
For electrons with a thermal injection spectrum, the model predicts
two types of HXR coronal sources.
(a) Thermal sources are formed in traps dominated by the betatron
mechanism.
(b) Non\-thermal sources with a power-law spectrum appear when
electrons are accelerated by the Fermi mechanism.
With account of rare Coulomb collisions inside the trap,
a double-power-low spectrum is formed from a power-law spectrum of
electron injection from SHTCL \citep{Bogachev09}.
Fermi acceleration has significant advantages in collapsing magnetic
traps as compared with the betatron mechanism which mainly heats
the low-energy electrons \citep[][Chapter~7]{Somov07}.
%
%%%%%%%%%%%%%%%%%%%%%%%%%%%%%%%%%%%%%%%%%%%%%%%%%%%%%%%%%%%%%%%%%%%%%%%%
%

It is important to note that the emission measure of emitting
fast electrons, trapped and accelerated inside the collapsing
loop, initially grows slowly with decrease of the loop
length $ L (t) $. As calculated by \cite{Bogachev07},
see their Fig. 3a,
the emission measure reaches its maximal value only when
the loop length becomes as small as
of about 0.2--0.1 of its initial length $ L (0) $.
With further shrinkage of the loop, the emission measure
decreases quickly to zero.
As a consequence, the flux of hard X-ray emission from the
collapsing trap attains a maximum at approximately the same
values of remaining length of the loop \citep[see
l (t) = L(t)/L(0) $\sim$ 0.2 in Fig. 4a in][]{Bogachev07}.
On the other hand, when the loop is just created by
reconnection process at the X-point inside a thin reconnecting
current layer \citep[e.g,][]{Ugai08},
it is strongly stressed by magnetic tensions along magnetic
field lines in direction from the X-point to the edge of the
current layer.
That is why the loop shrinks and becomes less stressed.
Moreover, the loop goes down less quickly with the progress
of time because the magnetic stress is decreasing
continuously.
In other words, the loop becomes less non-equilibrium, more
close to a state with minimal energy, i.e. the potential state
\citep{Somov06}.
In such a state, the height of the loop is naturally
proportional to the distance between its feet.
Therefore, it is reasonable to assume that, at the time when
the collapsing loop has the maximal brightness in hard X-rays,
the height of the LT coronal HXR sources is proportional to
the distance between the conjugate FP sources in the chromosphere.

%Anyway, at a certain length of a collapsing trap or (which is the
%same) at a certain height of its top above the photosphere, a given
%magnetic loop becomes to go down less quickly;
%its top becomes very bright in HXRs but fades in HXRs later on.
%Let us assume that such a height, at which the HXR brightness of a
%collapsing trap under consideration has a maximum, is proportional to
%the distance between the FPs of the loop at this instant.
%This means that the loop becomes less stressed in comparison with
%its initial state, when it was just created by reconnection at the
%X-point inside a thin reconnecting current layer \citep[e.g,][]{Ugai08}.
%{\bf In other words, the loops becomes less non-equilibrium and, therefore,
%more close to a potential state \citep{Somov06}}.

Now we shall discuss the consequences of the last assumption
coming back to the rainbow reconnection model.
%Now we shall consider the consequences of the last assumption
%coming back to the rainbow reconnection model.
Figure~\ref{flare_cartoon}a
%
%%% Fig. 15 (a) %%%
%
shows different flux tubes~$ f_{1} $, $ f_{2} $ etc reconnected at
different times~$ t_{1} $, $ t_{2} $ etc;
here $ t_{2} > t_{1} $ etc.
The first flux tube, the loop~$ f_{1} $ manifests two
FPs~$ P_{a} $ and $ P_{b} $ and a loop\-top HXR source~$ LT $.
Figure~\ref{flare_cartoon}a
%
%%% Fig. 15 (a) %%%
%
illustrates two effects.
The first one is the well-known classical effect, an increase of the
distance between the flare ribbons because of reconnection in the
coronal SHTCL.
The displacements~$ \delta x $ of FPs are antiparallel
to the converging components~$ {\bf V}_{\bot} $ of the velocity field
in the photosphere.
This means that two flare ribbons move out from the $ PNL $
as predicted by the standard model of two-ribbon flare.

%
%%%%%%%%%%%%%%%%%%%%%%%%%%%%%%%%%%%%%%%%%%%%%%%%%%%%%%%%%%%%%%%%%%%%%%%%
%
The second effect is less trivial.
The displacements~$ \delta y $ of the FPs are parallel to the PNL
because of relaxation of the non\-potential component of the field
created by the photospheric shear flows before a flare.
So this is the magnetic shear relaxation.
The displacements~$ \delta y $ of FPs are antiparallel
to the photospheric shear velocity~$ V_{\parallel} $.
Since the photospheric shear mainly dominates in the vicinity of
the $ PNL $, during the first phase of a flare
$ \delta y \gg \delta x $.

In order to interpret our observational results related to the
first phase, let us consider a magnetic loop~$ f_{1} $ which after the magnetic shear relaxation process evolves into loop~$f_{2}$. The magnetic loop~$ f_{1} $ has a larger altitude of the loop\-top HXR
source (LT) because the distance between its FPs, $ P_{a} $
and~$ P_{b} $, is larger than the distance between the FPs of loop $f_{2}$, $ P_{a}^{\, \prime} $ and $ P_{b}^{\, \prime} $.
% undergoes created earlier
%than the loop~$ f_{2} $ has a larger altitude of the loop\-top HXR
%source~$ LT $ because the distance between its FPs~$ P_{a} $
%and~$ P_{b} $ is larger than the distance between
%FPs~$ P_{a}^{\, \prime} $ and $ P_{b}^{\, \prime} $.
%This is why the coronal LT source is located at higher altitude in the %loop~$ f_{1} $ than that of the loop~$ f_{2} $
%is located higher in the corona then the loop\-top
%source in the tube~$ f_{2} $.
%This is why the coronal loop\-top source~$ LT $ in the loop~$ f_{1} $
%is located higher in the corona then the loop\-top
%source in the tube~$ f_{2} $.
As a result, an apparent motion of the coronal HXR source is
directed downward ($ \delta z < 0 $) and the two conjugate FPs sources converge. 
%Therefore, our model explains the descending motion of the coronal source and decrease in the FPs separation during the first phase of the event as a consequence of the magnetic shear relaxation process. 
The descending motion of the coronal source is typical for the first phase in the flare of October 24, 2003, as well as presumably in many other flares
\citep[see][]{Sui03,Sui04,Li05,Ji06,Veronig06,Ji07,Joshi07,Liu09}.
The observations of larger separation between the two conjugate X-ray FPs observed in the first phase than that in the second phase (cf. Figure~\ref{rhessi_25-50}) also seems to be consistent with our model.
However, the detailed analysis FPs motion during the first phase was not possible for this event as FPs were visible for a very short time. %during the first phase.  

% Many recent studies have presented the simultaneous observations of the %descending LT as well as converging FPs motions\citep{Ji06,Ji07} and %\cite{Liu09}.}
%Many recently studied flare events show that during the rising phase the %flaring loops or LT source have a descending motion and, at the same time, %flare ribbons or FPs are converging \citep{Ji06,Ji07,Liu09}.}

%\citep[see][]{Ji06,Ji07,Liu09}.}

%\citep[see][]{Sui03,Sui04,Li05,Ji06,Veronig06,Ji07,Joshi07,Liu09}.
%The simultaneous observations of converging motion of conjugate FPs and descending LT source have been reported recently by many authors \citep{Ji06,Ji07,Liu09}. 

%\cite{Ji06,Ji07} have reported the simultaneous observations of LT descending and footpoints converging motions. 

%
%%%%%%%%%%%%%%%%%%%%%%%%%%%%%%%%%%%%%%%%%%%%%%%%%%%%%%%%%%%%%%%%%%%%%%%%
%
The second phase of a flare involves reconnection of the magnetic
field lines (and flux tubes) whose FPs are located at
larger distance from the $ PNL $ and, therefore, have a smaller shear
or practically none as schematically illustrated by
Figure~\ref{flare_cartoon}b. Here the apparent displacements~$ \delta x $ of FPs are
directed away from each other and from the $ PNL $.
The distance between FPs becomes larger with time,
and the LT~HXR source moves upward in agreement with the
standard model.
This seems to be fairly true for eruptive flares in the later stage
when the primary energy source (i.e. reconnection site) is located high
enough in the corona.
It looks like that the standard model of the two-ribbon flares seems
to be ``asymptotically'' true at later stages.
However, this is not necessarily a charitable trust. Here it is noteworthy that same physical concepts concerning the photospheric motions may apply to other events such as the Bastille Day Flare of 2000 July 14, although more complicated models are required to explain the observational characteristics of such a large event \citep[]{Somov02,Somov05b}.

In 72 flare events analyzed by
So\-mov et al. (2005a)
on the basis of the {\em Yohkoh\/} {\em HXT\/} data,
it was not simple to distinguish a flare with a decreasing FP
separation parallel to the $ PNL $, as discussed above,
from a flare with increasing separation also parallel to the $ PNL $,
because both kinds of separation were present in the same flare.
In the onset of such a flare, the FP sources move toward each other and
the distance between them decreases.
Then they pass through a ``critical point''.
At this moment, the line connecting the sources is nearly
perpendicular to the straight $ PNL $;
in general, we should characterize the photospheric magnetic field by
a {\em smoothed, simplified\/} neutral line \citep[$SNL$, see][]{Gorbachev89}
which is not a straight line.
%At this moment, the line connecting the sources is nearly
%perpendicular to the $ PNL $;
%in general, we characterize the photospheric magnetic field by
%a {\em smoothed, simplified\/} neutral line
%\citep[$SNL$, see][]{Gorbachev89}.
After that moment, the FP sources move away from each other with increasing
separation between them \cite [for example, the M4.4 flare on 2000
October~29 at 01:46~UT;][see Figure~2] {Somov05}.
Such a motion pattern seems to be similar to that one predicted by the
rainbow reconnection model after the critical moment shown in
Figure~\ref{flare_cartoon}b.
%
%%% Fig. 15 (b) %%%
%
Starting that moment, the shear relaxation must continue
($ \delta y > 0 $)
if the excess of coronal magnetic tensions have not been completely released during 
the first phase. Otherwise the ordinary reconnection process ($ \delta y = 0 $)
dominates during the second phase describing the ordinary energy
release in terms of the classical standard model.
Note that in both cases $ \delta x > 0 $ and, therefore,
during the second phase
the FP separation increases
and, as a consequence, the height of the loop\-top HXR sources
increases too.

In summary, recent multi-wavelength studies of solar flares have revealed a kind of two evolutionary phases of solar flares in terms of the motion of LT source. In the initial phase (or the first phase) the LT source exhibits a downward motion for a few to several minutes and the explanation of this phase is beyond the scope of classical CHSKP scenario of eruptive flares. The analysis of 2003 October 29 M7.6 flare, presented in this paper, provides one of the clearest observations of the descending LT source with X-ray and EUV images. Recent studies have revealed another important aspect of the first phase in which the descending LT source motion in temporally associated with the converging motion of FPs sources \citep{Ji06,Ji07,Liu09} which is consistent with the rainbow reconnection model discussed in this paper. 
%In the present event, however, it was not possible to make a detail analysis %FPs motion during the first phase as were visible for a very short time %during the first phase.    
\cite{Ji07} suggested another explanation of the first phase which is based on magnetic implosion conjecture \citep{Hudson00} that predicts contraction of coronal field lines simultaneously with the energy release. In the framework of sheared linear force-free arcade, \cite{Ji07} showed that the release of magnetic energy will reduce magnetic shear of the arcades and that the less sheared arcades will have smaller height and span.
\cite{Liu09} further found that the double FPs first do move toward and then away from each other, mainly
parallel and perpendicular to the $ PNL $, respectively, and that the transition from the first to the second type of
these apparent FP motions coincides with the direction
reversal of the motion of the loop\-top source. 
%Therefore an important aspect of future investigations would be to statistically examine the relationship between the converging motion of FPs sources and the descending LT source motion.
Therefore an important aspect of future investigations would be to examine the relationship between the converging motion of FPs sources and the descending LT source motion for more solar flares using existing data and new observations.

\acknowledgments
We acknowledge RHESSI and TRACE for their open data policy.
RHESSI and TRACE are NASA's small explorer missions. The H$\alpha$ data used in this paper was obtained from Solar Tower Telescope at Aryabhatta Research Institute of observational sciencES (ARIES), Nainital, India. This work has been supported by the ``Space Weather Center Project" 
of KASI and KASI basic research fund.
This work was partially supported by the Russian Foundation for Fundamental Research (Project No. 0802-01033-a). A.M.V. acknowledges support of the Austrian Fonds zur F\"orderung der wissenschaftlichen Forschung (FWF grant P20867-N16). J.L. was supported by NSF grant AST-0908344. Y.J.M has been supported by the WCU grant (No. R31-10016) funded by the Korean Ministry of Education, Science and Technology and by the Korea Research Foundation Grant funded by the Korean Government (MOEHRD, Basic Research Promotion Fund)
(KRF-2008-314-C00158 and 20090071744). B.J. thanks Wei Liu for useful discussions on RHESSI data analysis. We sincerely acknowledge the constructive comments and suggestions of the referee of the paper.

%\bibliographystyle{apj}
%\bibliography{bhuwan}

\begin{figure*}
\centering
\includegraphics[width=\textwidth]{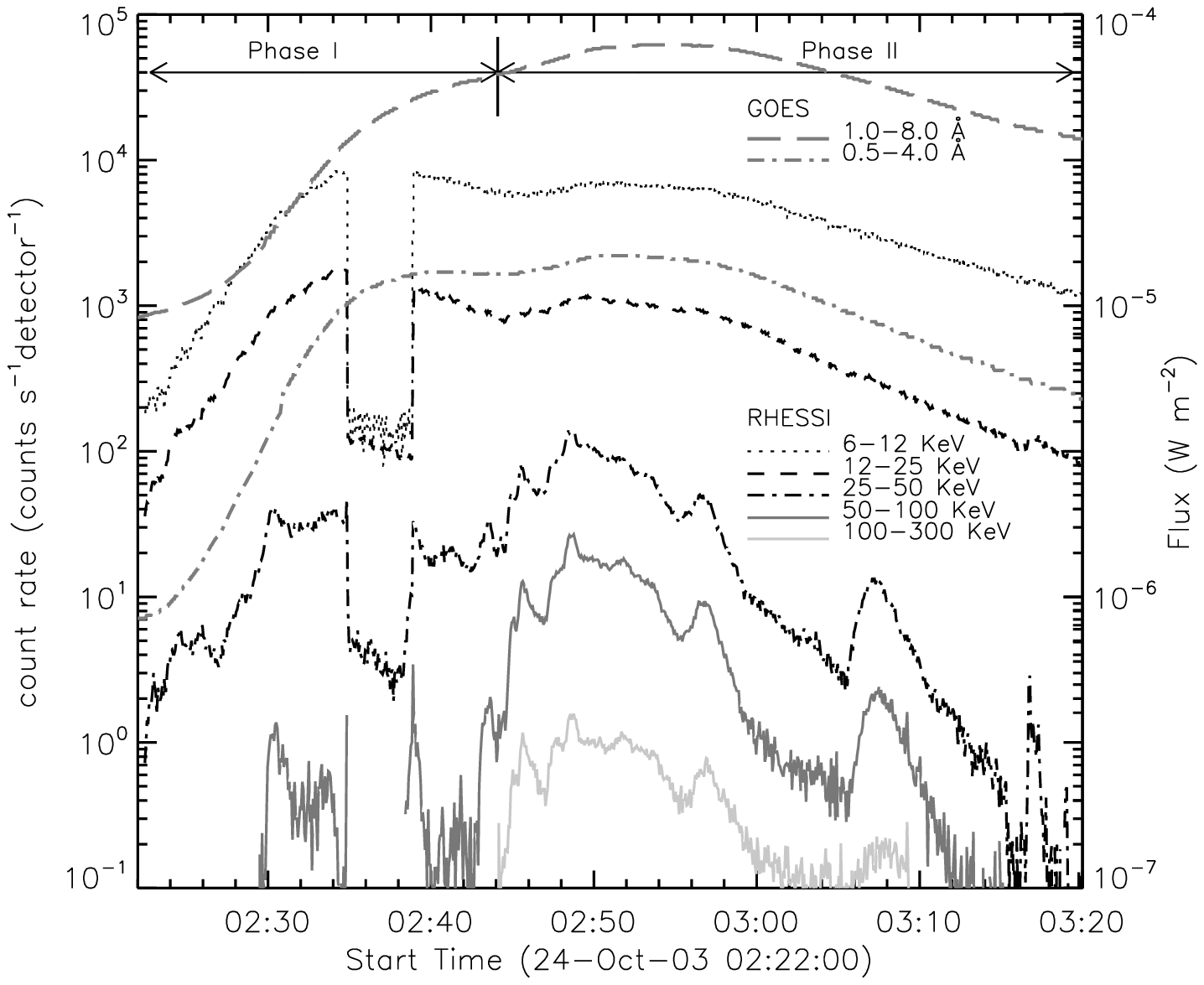}
\caption{RHESSI and GOES light curves of the flare. RHESSI counts rates are averaged over every 4 sec. In order to present different RHESSI light curves with clarity, the RHESSI count rates are scaled by factors of 1, 1/4, 1/5, 1/10, and 1/50 for the energy bands 6--12, 12--25, 25--50, 50--100, and 100--300 keV, respectively.}
\label{goes_rhessi}
\end{figure*}

\begin{figure*}
\epsscale{.80}
\includegraphics[width=\textwidth]{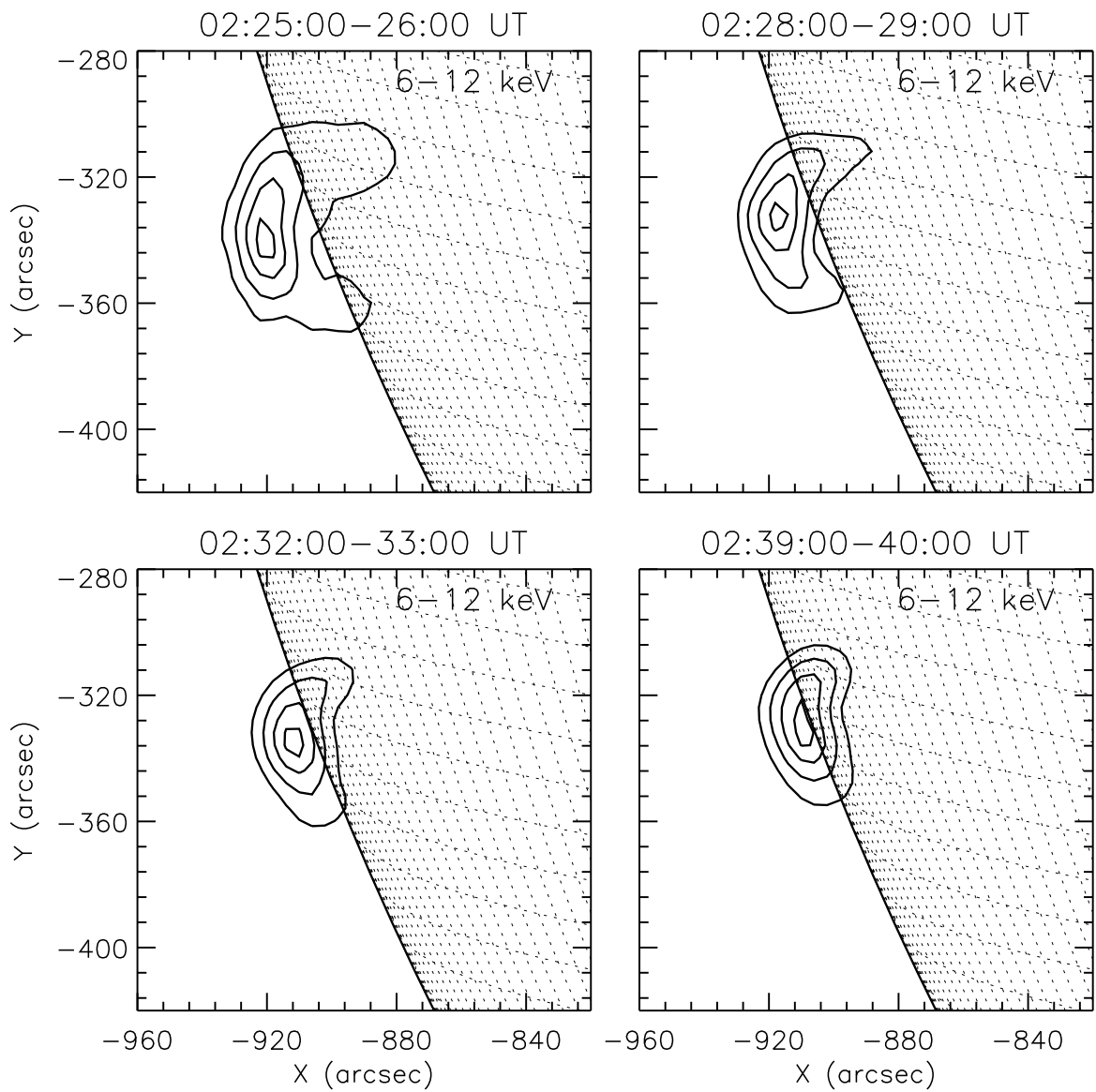}
\caption{Sequence of RHESSI 6--12 keV images reconstructed with the CLEAN algorithm using grids 3 to 8 and natural weighing scheme
during the first phase of the flare showing the altitude decrease of the flaring loop system.  The integration
time for each image was 1 min. The contour levels are 40\%, 60\%, 80\%, and 95\% of the peak flux in each image. The solar limb and
heliographic grids are also drawn in each panel.}
\label{rhessi_first_ph}
\end{figure*}

\begin{figure*}
%\epsscale{.80}
\includegraphics[width=\textwidth]{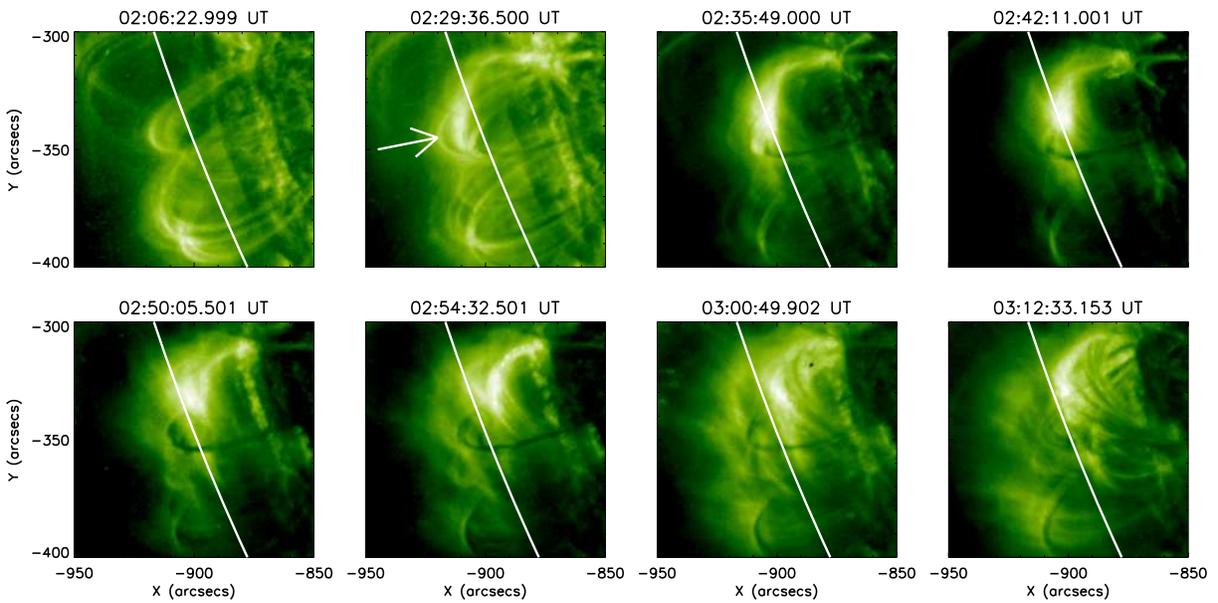}
%\plotone{rhessi_plot.ps}
\caption{Time sequence of TRACE 195~{\AA} images showing a closer view of coronal loops associated with the flaring region. The bright
region at the top of the loop system is indicated by an arrow in the image at 02:29:36 UT. 
Note the pre-existing loop system and altitude
decrease of the flaring loops at the first flare phase, followed by ($\gtrsim$ 2:54~UT)
post-flare loops appearing dark and structured.} 
%(1.5 MK TRACE 195~{\AA} response).
\label{trace195_series}
\end{figure*}

\begin{figure*}
%\epsscale{.80}
\includegraphics[width=\textwidth]{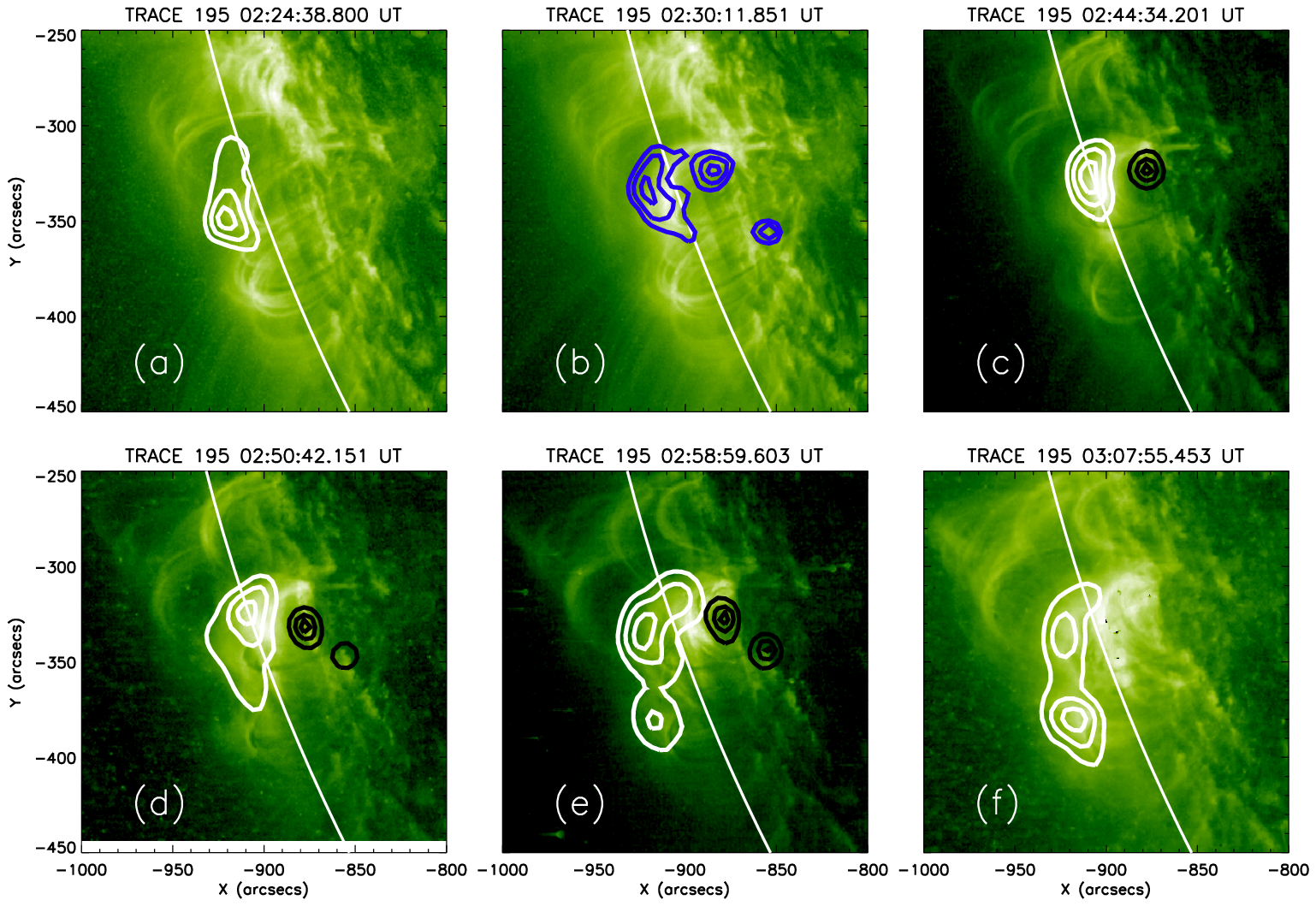}
%\plotone{rhessi_plot.ps}
\caption{Sequence of TRACE 195~{\AA} images overlayed by co-temporal RHESSI X-ray images. RHESSI image parameters are the same as in Fig. \ref{rhessi_first_ph}. 
Panels (a)-(f) (except b): 6--12 keV (white contours) and 50--100 keV (black contours) images are plotted with contour levels 50\%, 70\%, and 90\% of the peak flux in each image. Panel (b): 25--50 keV image (blue contours) is plotted with contour levels 65\%, 80\%, and 95\% of the peak flux.}
\label{trace195_hessi}
\end{figure*}

\begin{figure*}
%\epsscale{.80}
\includegraphics[width=\textwidth]{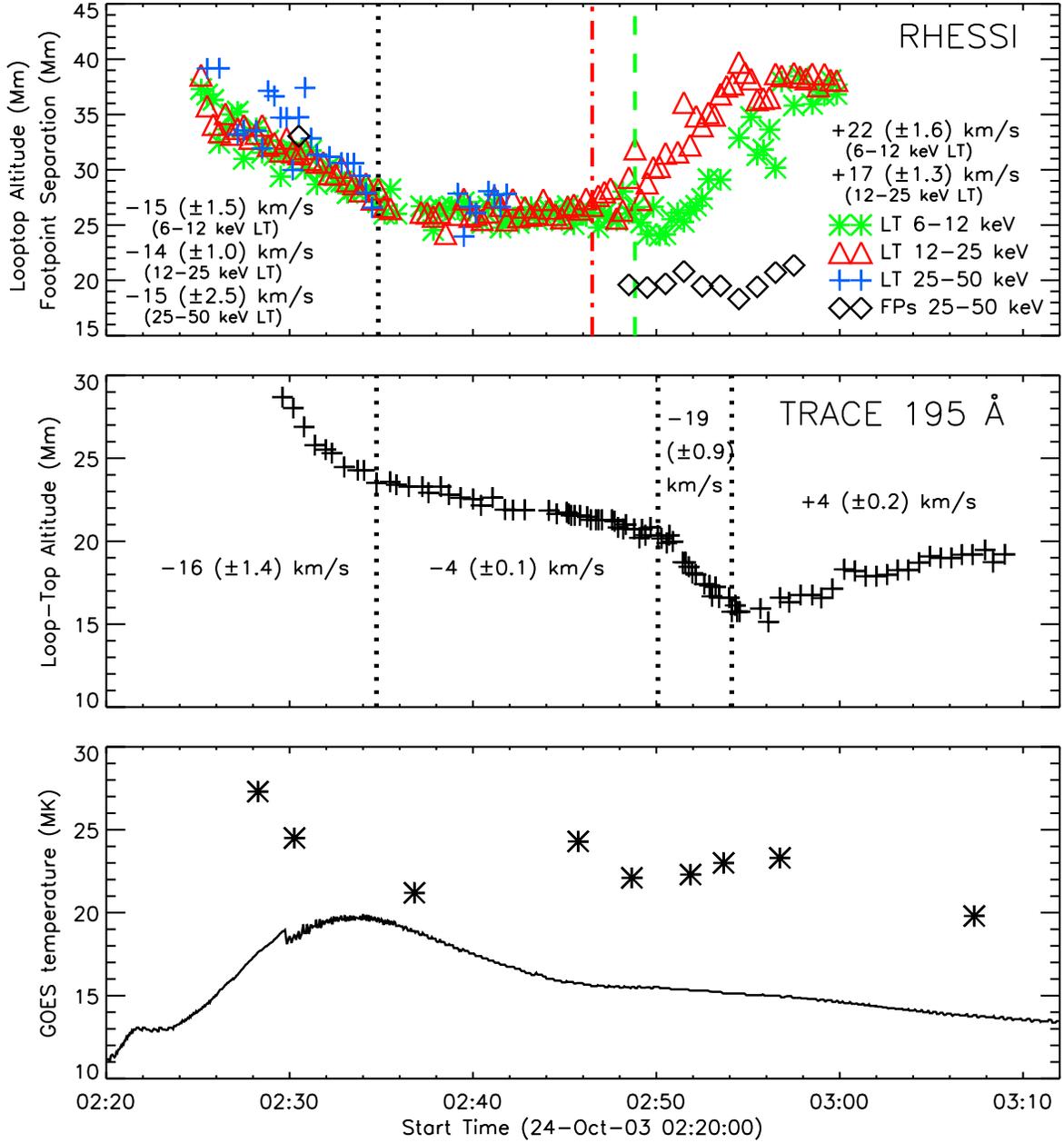}
%\plotone{rhessi_plot.ps}
\caption{Evolution of the altitude of the RHESSI LT source observed in the 6--12, 12--25, and 25--50 keV energy bands. Note that the RHESSI
LT source at 12--25 and 6--12 keV show upward motion after $\sim$02:46 and $\sim$02:49 UT respectively (start time is indicated by
dash-dotted and dashed lines respectively). The FP separation derived from RHESSI 25--50 keV images is also plotted. Middle panel: Evolution of the altitude of the TRACE 195~{\AA} LT source. The mean velocities
derived by linear fits to the altitude data in certain time interval (indicated by vertical lines) are annotated in the Figure. Bottom panel: Time profile
of temperature derived from GOES-12 measurements. The data points denoted by asterisk symbols show the plasma temperature estimated from RHESSI measurements.}
\label{trace_rhessi_LT}
\end{figure*}

\begin{figure*}
%\epsscale{.80}
\includegraphics[width=\textwidth]{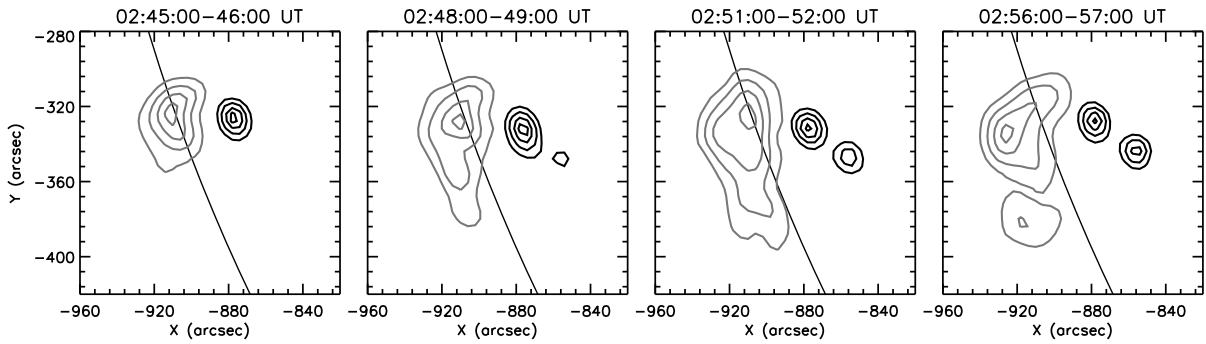}
%\plotone{rhessi_plot.ps}
\caption{Sequence of RHESSI 10-15 keV (gray contours) and 50-100 keV (black contours) images showing the evolution of LT and FP sources respectively. The image reconstruction parameters are the same as in Fig. \ref{rhessi_first_ph}. The contour levels are 40\%, 60\%, 80\%, and 95\%
of the peak flux in each image.}
\label{rhessi_second_ph}
\end{figure*}

\begin{figure*}
\includegraphics[width=\textwidth]{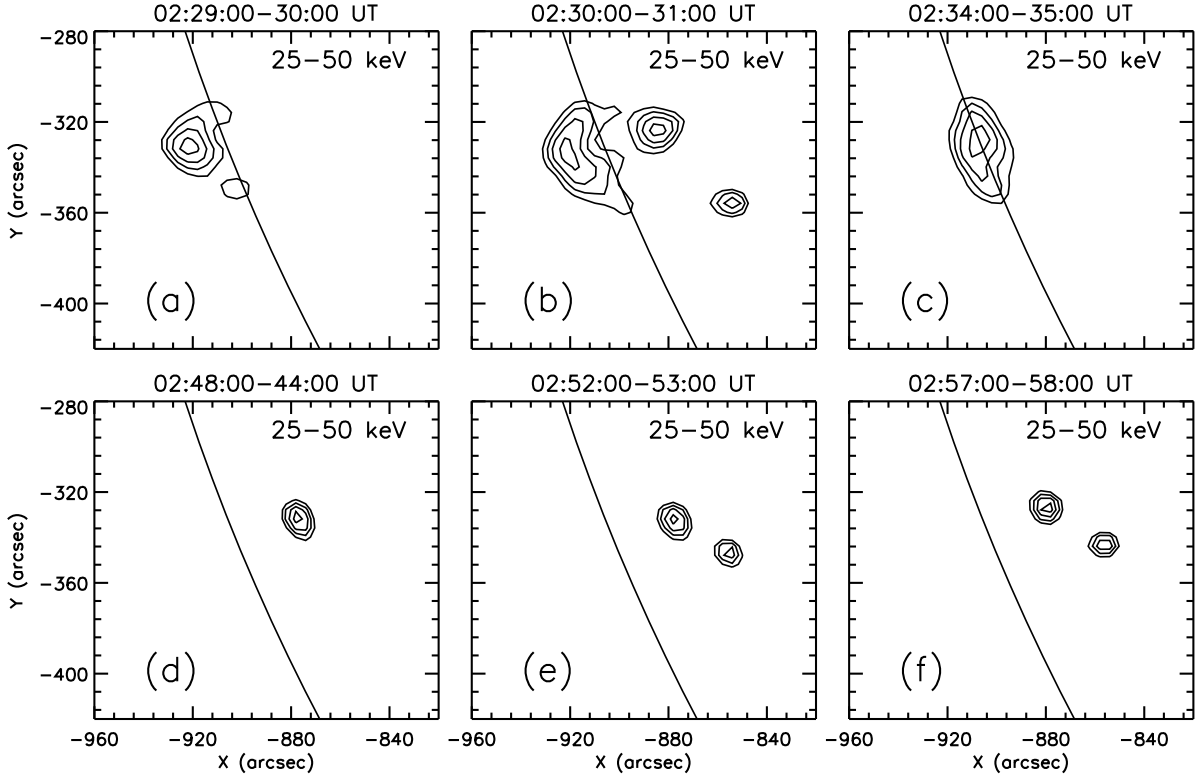}
\caption{Sequence of RHESSI 25-50 keV images. The contour levels are 65\%, 75\%, 85\%, and 95\% of the peak flux in each image. Note the two bright FP sources at 02:30:00-02:31:00 UT appeared for a short period during the prolonged preheating phase when the LT source shows a descending motion. On the other hand, two conjugate FPs can been seen during the second phase from 02:52 UT onward (see panel e). In this phase, the LT source moves upward (see Figure~\ref{rhessi_second_ph}). It is noteworthy that the separation between the two FPs during the first phase (panel b) is larger than that during the second phase (panel e).}
\label{rhessi_25-50}
\end{figure*}

\begin{figure*}
%\epsscale{.80}
\includegraphics[width=\textwidth]{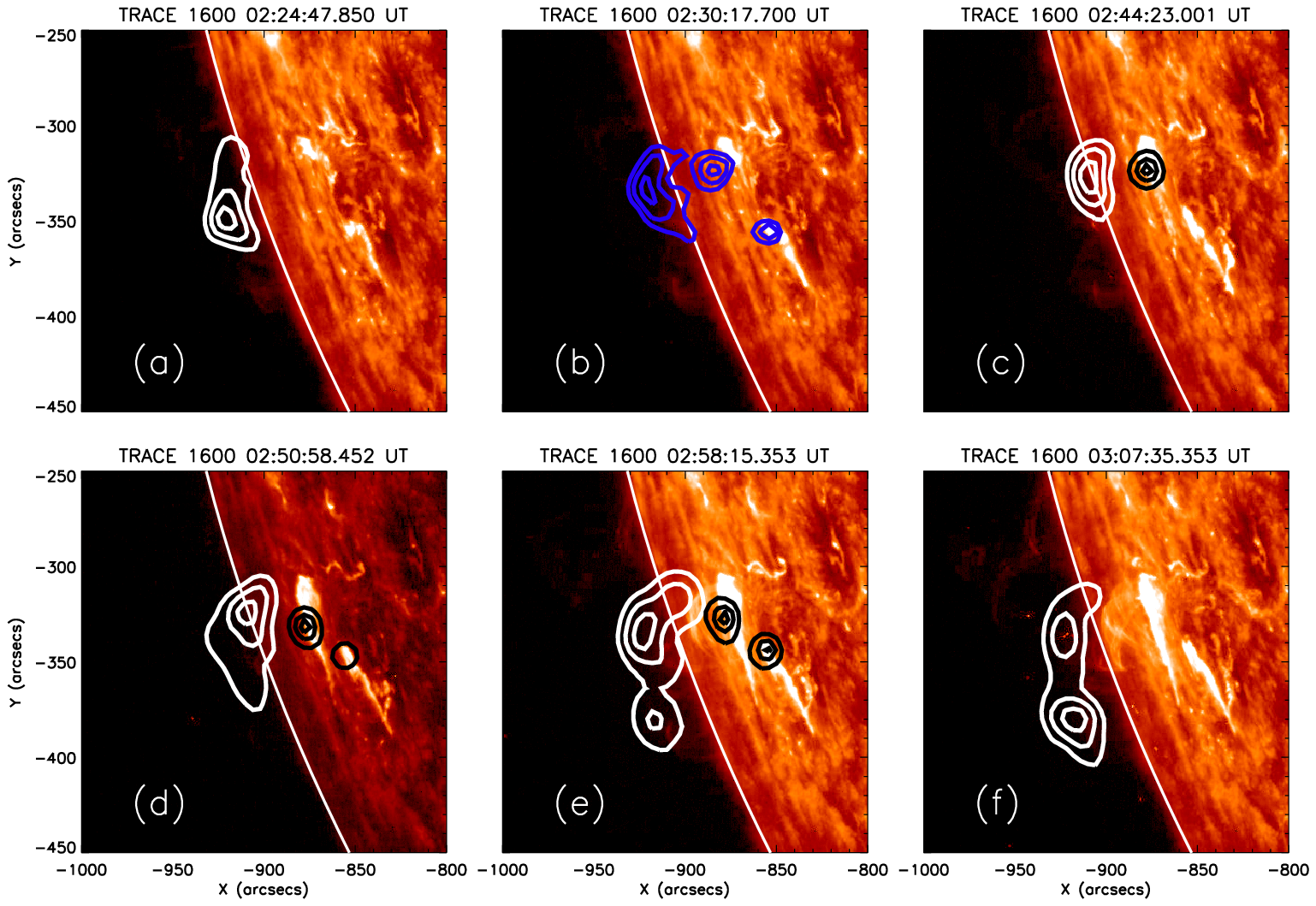}
%\plotone{rhessi_plot.ps}
\caption{Sequence of TRACE 1600~{\AA} images overlayed by co-temporal RHESSI X-ray images. RHESSI image parameters are the same as in Fig. \ref{rhessi_first_ph}. Panels (a)-(f) (except b): 6--12 keV (white contours) and 50--100 keV (black contours) images are plotted with contour levels 50\%, 70\%, and 90\% of the peak flux in each image. Panel (b): 25--50 keV image (blue contours) is plotted with contour levels 65\%, 80\%, and 95\% of the peak flux.}
\label{trace1600_hessi}
\end{figure*}

\begin{figure}
\epsscale{.80}
\plotone{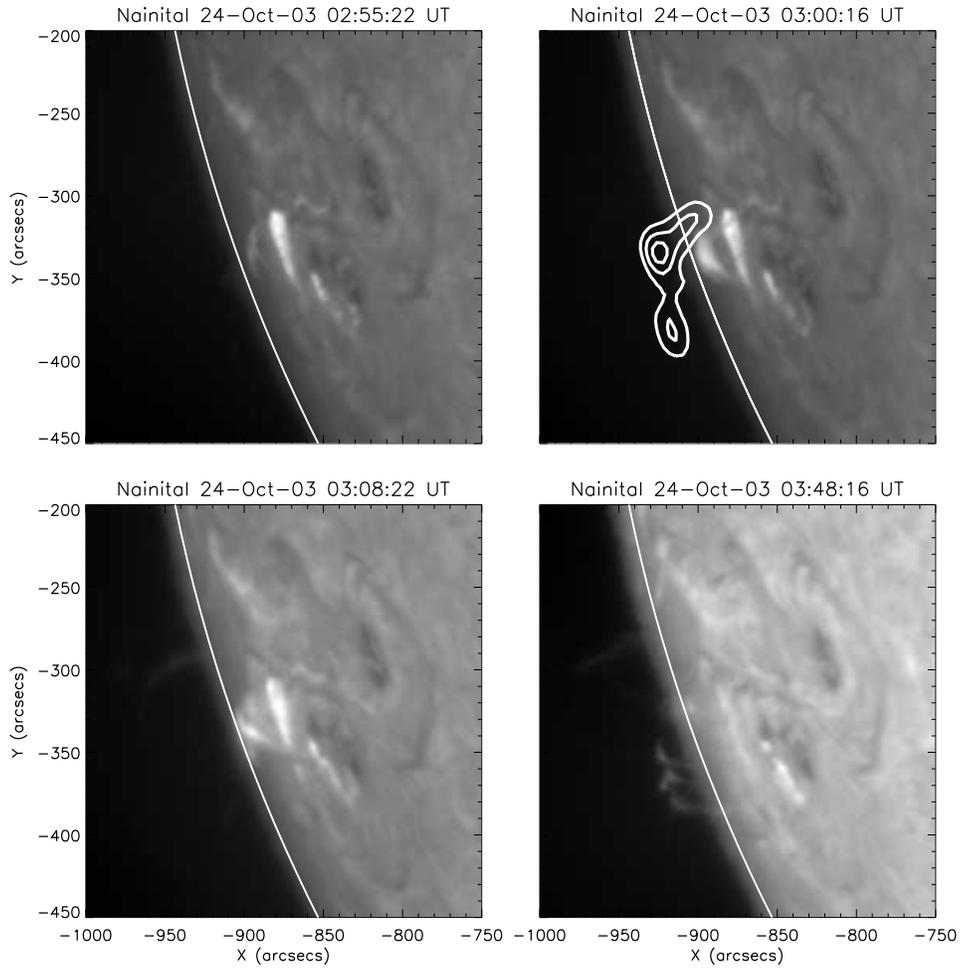}
\caption{Some representative H$\alpha$ images during the decline phase of the event. The white contours (only in top right panel) are RHESSI 
6--12 keV image. The contour levels are 50\%, 70\%, and 90\% of the peak flux. Note that the H$\alpha$ post-flare loops appearing bright
against the disk, implying high plasma densities in the loops.}
\label{halpha_hessi}
\end{figure}

%\begin{figure}
%\epsscale{0.6}
%\plotone{f11.eps}
%\caption{RHESSI light curves in three energy bands: 10--25, 25--100, and 100--300~keV. The time intervals during which the spectra shown in Fig.~\ref{rhessi_spec} were accumulated are indicated by vertical lines. Note that during 02:35--02:39~UT the RHESSI thick attenuator was in, whereas during all other times only the thin shutter was in. 
%}
%%\label{rhessi_ts}
%\end{figure}

\begin{figure}
\epsscale{0.7}
\plotone{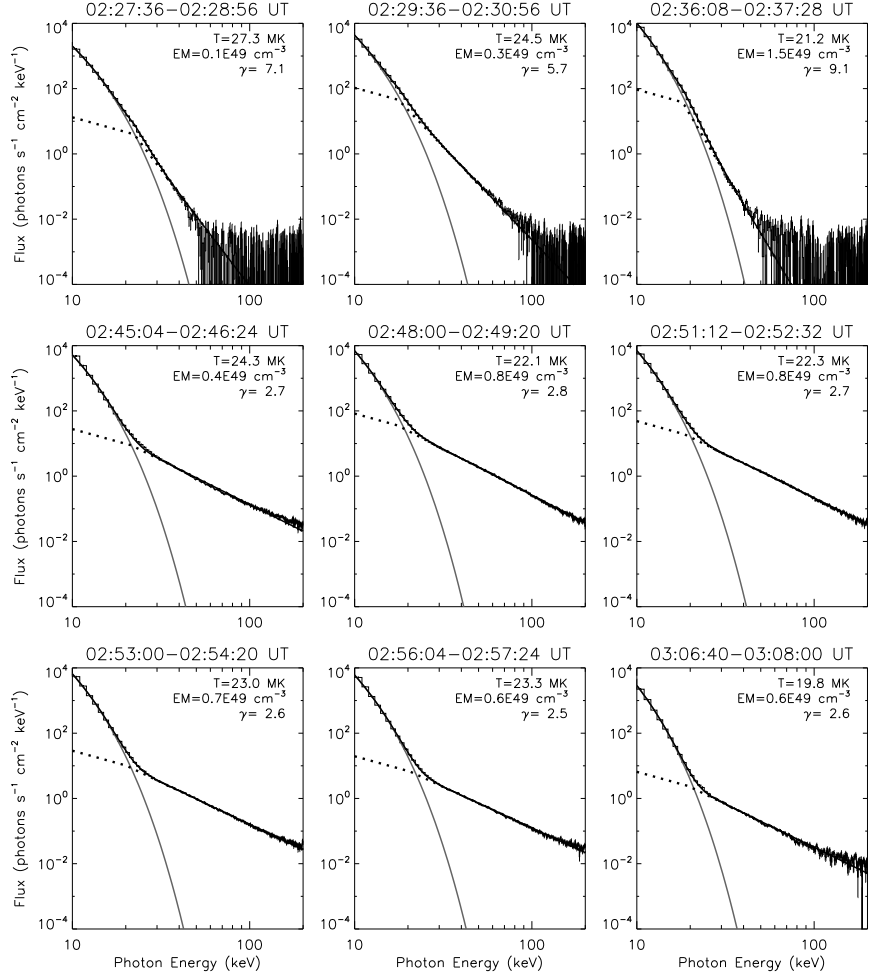}
\caption{Spatially integrated background-subtracted RHESSI spectra derived during nine time intervals during the flare together with the applied spectral fits. 
%The spectra in panels a)--c) were accumulated over 80~s during the first %phase of the flare, panels d)--i) during the second phase of the flare. 
The spectra were fitted with a thermal bremsstrahlung model (grey lines) and a functional power-law with a turn-over at low energies (dotted lines). The black full lines indicate the sum of both components. The early spectra (top row) were fitted in the energy range 10--50~keV, the spectra derived during the second phase of the flare (middle and bottom rows) were fitted in the range 10--200 keV. 
%Note that during time interval the RHESSI thick attenuator was in (A3 %state), whereas during all other intervals only the thin attenuator was in %(A1 state).
}
\label{rhessi_spec}
\end{figure}

\begin{figure*}
\epsscale{0.7}
\plotone{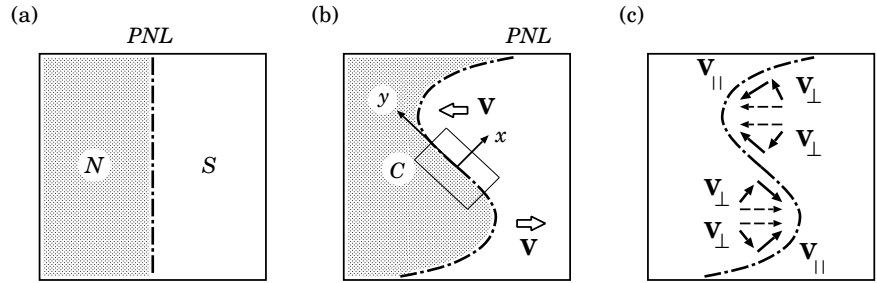}
\caption{(a) Model distribution of a vertical component of magnetic
             field in the photospheric plane.
         (b) A large-scale vortex flow in the photosphere distorts
             the photospheric neutral line $ PNL $.
         (c) A schematic decomposition of the velocity field
             $ {\bf V} $ into the components parallel and
             perpendicular to the $ PNL $.}

\label{mag_photo}
\end{figure*}

\begin{figure*}
\epsscale{0.7}
\plotone{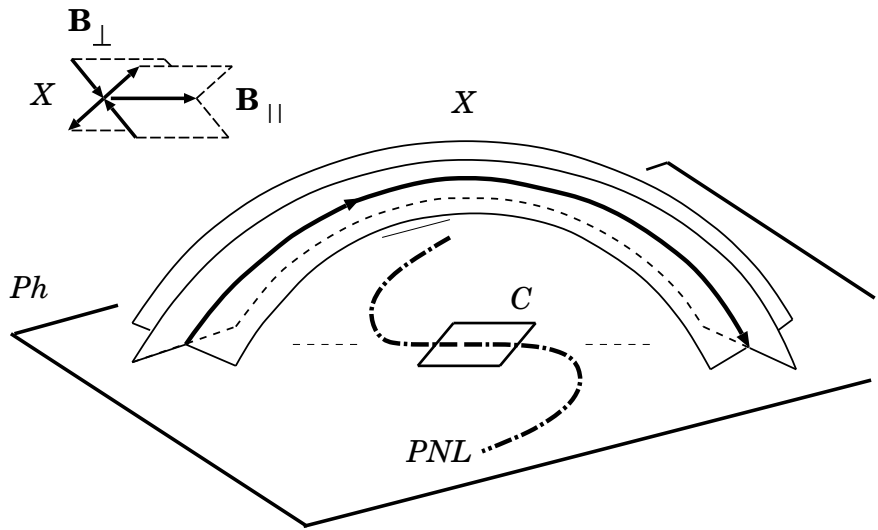}
\caption {The ``rainbow reconnection'' model
            (Somov 1985).
            Separator $ X $ above the $ S $-shaped bend of the
            $ PNL $.
            The inset in the upper left-hand corner shows the
            structure of magnetic field near the top of the
            separator with a longitudinal
            component~$ {\bf B}_{\parallel} $. }
\label{rainbow}
\end{figure*}

\begin{figure*}
\epsscale{0.5}
\plotone{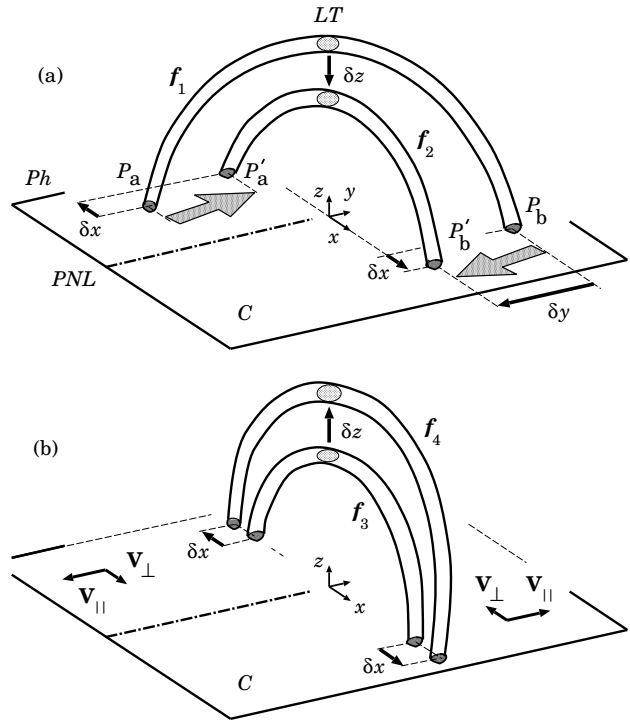}
\caption {Schematic representation of the flare evolution in terms
            of the rainbow reconnection model.
   (a) The first phase: Rapidly decreasing FP separation
       dominates an increase of distance between flare
       ribbons.
       The LT~HXR source goes down.
   (b) The second phase: The bright HXR kernels separate in
       opposite directions from the $ PNL $ and from each other.
       The LT moves upward.}
\label{flare_cartoon}
\end{figure*}

\end{document}